\documentclass[prl,twocolumn,showpacs,preprintnumbers,amsmath,amssymb]{revtex4}

\usepackage{graphicx}
\usepackage{amsmath}
\usepackage{dcolumn}
\usepackage{bm}

\begin{document}

\title{Direct Determination of Electron-Phonon Coupling Matrix Element in a Correlated System}

\author{Huajun Qin$^1$, Junren Shi$^1$, Yanwei Cao$^1$, Kehui Wu$^1$, Jiandi Zhang$^2$, E. W. Plummer$^2$, J. Wen$^3$, Z. J. Xu$^3$, G. D. Gu$^3$, and Jiandong Guo$^1$}

 \affiliation{$^1$Beijing National Laboratory for Condensed-Matter Physics \& Institute of Physics, Chinese Academy of Sciences, Beijing 100190, China}
 \affiliation{$^2$Department of Physics and Astronomy, Louisiana State University, Baton Rouge, LA 70808, USA}
 \affiliation{$^3$Brookhaven National Laboratory, Upton, NY 11973, USA}

\date{\today}

\begin{abstract}
High-resolution electron energy loss spectroscopy measurements have been carried out on an optimally doped cuprate Bi$_{2}$Sr$_{2}$CaCu$_{2}$O$_{8+\delta}$. The momentum-dependent  linewidth and the dispersion of an $A_1$ optical phonon are obtained. Based on these data as well as the detailed knowledge of the electronic structure from angle-resolved photoemission spectroscopy, we develop a scheme to determine the full structure of electron-phonon coupling for a specific phonon mode, thus providing a general method for directly resolving the EPC matrix element in systems with anisotropic electronic structures.
\end{abstract}


\maketitle

The interaction between electrons and various collective excitations (bosons) is the central ingredient for understanding many of the novel physical properties in condensed matter systems~\cite{RMP62-1027,PR338-1,JPCM20-043201}. For simple systems with isotropic electronic structures, such an interaction (electron-boson coupling, EBC) can be characterized by the Eliashberg spectral function $\alpha^2F(\omega)$~\cite{Grimvall}, which describes the energy of the bosonic modes involved in the coupling as well as their coupling strengths. Direct experimental determination of the Eliashberg function has proven to be important. For instance, in Pb, it played a crucial role in establishing the BCS theory of superconductivity~\cite{JMRowell}. Experimentally, there exists several techniques for determining the Eliashberg function, such as the McMillan-Rowell inversion method on the tunneling data of the conventional superconductors~\cite{JMRowell,McMillan}, and more recently, the maximum entropy method (MEM) in analyzing the quasi-particle dispersion kink observed in high-resolution angle-resolved photoemission spectroscopy (ARPES) measurements~\cite{JunrenMEM, XJZhouMEM, NonBi2201}. These approaches attempt to elucidate the EBC by probing its effects on the electrons. On the other hand, the effects of the EBC on bosonic modes is not commonly investigated.

For more complex systems, neither the isotropic Eliashberg function $\alpha^2 F(\omega)$ nor the anisotropic Eliashberg function $\alpha^2 F(\omega,\hat{\bm{k}})$ determined from ARPES is sufficient for fully characterizing EBC. A notable example is the high-$T_c$ cuprate, who has highly anisotropic electronic structure and the unconventional $d_{x^2-y^2}$ superconductivity. By probing the electrons alone, one can at the best determine the $\alpha^2 F(\omega,\hat{\bm{k}})$ with the dependence on the direction $\hat{\bm{k}}$ alone using ARPES and the aforementioned MEM analysis~\cite{Chien2009}. However, unlike the case of the conventional $s$-wave superconductors, the so-determined Eliashberg function cannot be directly related to the strength of the unconventional $d_{x^2-y^2}$ pairing.  This is because the Eliashberg function determined from the quasi-particle dispersion and the one dictating the $d_{x^2-y^2}$ pairing strength belong to different symmetries~\cite{Bulut1996}.  For these systems, it is necessary to resolve the full structure of the EBC, i.e., the matrix element $g(\bm{k}, \bm{k}^\prime)$, which characterizes the probability amplitude for the electron transition from $\bm{k}$ to $\bm{k}^\prime$ induced by the interaction with bosons~\cite{PRL93-117003,PRL93-117004}.  Because the electronic measurements such as ARPES only contain the information of the electron final state $\bm{k}^\prime$, with the contributions from the different initial states $\bm{k}$ integrated, it is in general impossible to determine the full structure of $g(\bm{k}, \bm{k}^\prime)$ by probing electrons alone. The manifestation of the EBC on the bosonic modes needs to be probed.

In this paper, we develop a method to resolve the full structure of EBC by probing the bosonic modes with high resolution electron energy loss spectroscopy (HREELS), in combination with the existing ARPES data for the electronic properties. The momentum-dependent linewidth as well as the dispersion of an $A_1$ optical phonon was recorded as a function of the direction relative to the crystalline orientation. The initial and final states involved in the coupling and the corresponding strength are simultaneously determined. The developed scheme is general and applicable to many systems with anisotropic electronic structures.

We have used a prototype high-temperature superconductor (HTSC) Bi$_{2}$Sr$_{2}$CaCu$_{2}$O$_{8+\delta}$ (referred to as BSCCO in the following)~\cite{PSS242-11,CampuzanoPRL83} as a test system. The electron band structure, Fermi surface, and surface properties have been well characterized~\cite{RMP75-473}. Many experimental studies have revealed the signature of the interaction between electrons and a collective bosonic mode. ARPES shows the electron self-energy renormalization in the form of kinks in the dispersion at the energy scales of $\sim$70 meV below the Fermi energy in the nodal region~\cite{N412-510,PRL85-2581,PRL87-177007,PRL96-017005} and $\sim$40 meV near the antinodal region~\cite{PRB68-174520,PRL93-117003}. Direct measurements of phonon spectra have been carried out with Raman spectroscopy~\cite{PRB53-11796}, infrared spectroscopy (IR)~\cite{PRB69-054511}, HREELS~\cite{PRB48-12936}, inelastic neutron scattering (INS)~\cite{PSS242-30} and inelastic x-ray scattering (IXS)~\cite{PRL100-227002} for BSCCO and other cuprates HTSCs. However, a systematic study of the momentum-dependent phonon renormalization is still missing and it is needed to elucidate the details of EBC.

The experiments were carried out in a commercial system containing variable temperature scanning tunneling microscope (STM) and HREELS (LK-5000). The base pressure was better than 1$\times$10$^{-10}$ mbar. The superconducting property of the optimal doped BSCCO single crystal was carefully characterized with the transition temperature at T$_{c}$ = 91 K. The sample was cleaved \textit{in situ} in vacuum and the surface was characterized with STM and low energy electron diffraction (LEED). Fig.~1~(a) shows that the cleaved surface was atomically flat with the Bi-O plane exposed. The cleavage quality was further verified by the presence of the (1$\times$5) superstructure that was visible in both the LEED patterns and STM images, which also indicates the relative angle of the crystalline orientation to the measurements geometry. The HREELS measurements were taken with the electron incident angle of 65$^{o}$, and excitation energy varied from 3.5 eV to 50 eV to achieve different detecting depth. All the observed spectral features showed no excitation energy dependence for a fixed momentum transfer parallel to the surface. Here we report the results taken with the excitation energy of 50 eV offering a wide momentum transfer range. The HREELS sample stage was coupled to a constant-flow liquid helium cryostat. And the measurements were taken at different temperatures from 60 to 300 K, across the superconducting T$_{c}$.

\begin{figure}
\includegraphics[clip,width=3.4in]{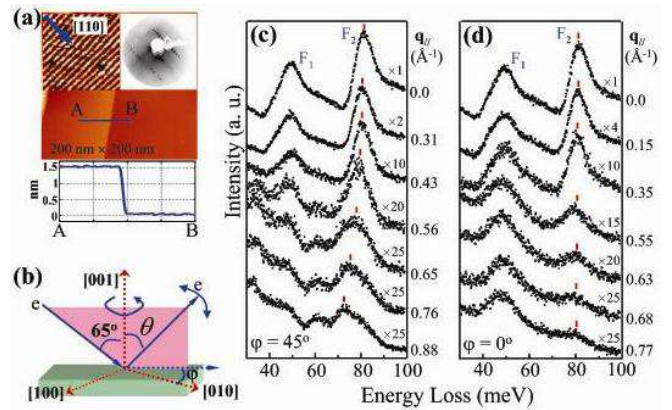}
\caption{(color online) (a) STM image of the cleaved BSCCO surface. The left inset shows the 1$\times$5 superstructure with the \textquotedblleft $\times$5\textquotedblright direction aligning to [110] orientation. The right inset shows the LEED patterns taken with beam energy of 61 eV at room temperature. The line profile along AB is displayed in the lower panel, which indicates a step of 1.5 nm corresponding to the interspacing between two cleavable BiO planes. (b) Schematic drawing of the scattering geometry for the angle-resolved HREELS measurements. (c) and (d) The HREELS spectra measured with $\varphi$ = 45$^{o}$ and 0$^{o}$, respectively, with the excitation energy of 50 eV at 60 K. The bars (red) are guides to the eye for the phonon energy shift.}
\end{figure}

Angle-resolved HREELS of the optimal doped BSCCO sample were measured along different directions relative to the crystalline orientation, as shown in Fig.~1. At Brillouin zone (BZ) center (the in-plane momentum transfer $\bm{q}$ = 0), two main features are resolved near 50 meV (F$_{1}$) and 80 meV (F$_{2}$), respectively. Feature F$_{1}$ corresponds to the out-of-plane vibration of oxygen atoms in Bi-O plane \cite{PRB53-11796,PRB69-054511,PRB48-12936}, with two shoulders at both sides with energy of $\sim$41 meV and $\sim$62 meV that have been reported earlier \cite{PRB48-12936}. Feature F$_{2}$, to be focused on in the following, appears as a main peak at $\sim$80 meV with a broad tail centered at $\sim$88 meV. The vibration along c axis of the apical oxygen atoms in the Cu-O semi-octahedra is responsible for the main peak \cite{PRB53-11796,PRB69-054511,PRB48-12936}, while the high-energy tail might be related to the disorder \cite{PRB53-11796}. Also considering the selection rule of HREELS, we are able to rule out other energy loss mechanisms that fall into this energy range, such as the phonons of in-plane breathing modes of oxygen in Cu-O$_{2}$ plane since they are not dipole active along the surface normal and therefore undetectable by HREELS at BZ center. Along any measurement direction, we detect no obvious temperature dependence across superconducting T$_{c}$.

As shown in Fig.~1~(c), the most prominent characteristic of the momentum-resolved HREELS presents along nodal direction ($\varphi$ = 45$^o$) where the energy dispersion of the 80-meV feature F$_{2}$ softens significantly from BZ center towards the boundary. Such a behavior has been quantitatively analyzed \cite{note1}. The momentum-dependent phonon energy and linewidth (full-width at half maximum, FWHM) along different directions relative to the crystalline orientation are shown in Fig.~2. An abrupt decrease from 81 meV to 74 meV occurs to the phonon energy with an onset at q$_{\parallel} \sim$ 0.45 \AA $^{-1}$ (0.54 $\pi$/a with a = 3.82 \AA). With the same onset, the FWHM increases suddenly from 8 meV at q$_{\parallel} \sim$ 0.45 \AA $^{-1}$ to 18 meV at q$_{\parallel} \sim$ 0.8 \AA$^{-1}$, indicating the intrinsic strong interaction between the apical oxygen phonon and electrons in BSCCO.

\begin{figure}
\includegraphics[width=3.3in,clip]{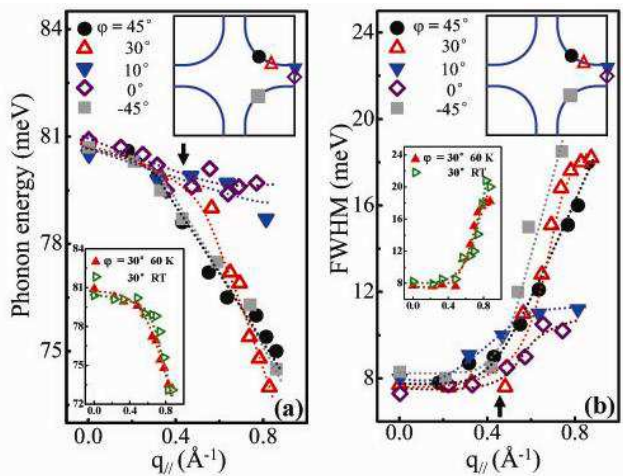}
\caption{(color online) (a) The energy dispersion of the apical oxygen phonon at 60 K. The different azimuth angels of q$_{\parallel}$ are illustrated in the upper inset in k space relative to the Fermi surface. When $\varphi > $10$^{o}$, there is a sudden drop with the onset of q$_{\parallel}\sim$ 0.45 \AA$^{-1}$. The lower inset compares data taken at different temperatures, indicating the same energy softening across superconducting T$_{c}$. (b) The dependence of linewidth on q$_{\parallel}$ along different azimuth angels. There is a sudden broadening when $\varphi > $10$^{o}$, with the onset aligning to that of energy dispersion at q$_{\parallel}\sim$ 0.45 \AA$^{-1}$, as indicated with the black arrows. Similarly, the lower inset shows the unchanged linewidth dependence on q$_{\parallel}$ across superconducting T$_{c}$. All the dotted lines are guides to the eye. Due to the existence of the (1$\times$5) superstructure, there are two irreducible nodal directions ($\varphi$ = $\pm$45$^{o}$). Our measurements show identical momentum-dependence along both directions.}
\end{figure}

The phonon dispersion and linewidth data show dramatic dependence on the measurement direction with respect to crystalline orientation ($\varphi$). As $\varphi$ decreases from 45$^{o}$ (the nodal region), the anomalous behavior persists until $\varphi$ gets close to the antinodal region ($\varphi \leq$ 10$^{o}$), when neither the energy nor the linewidth shows prominent momentum-dependence at temperatures above or below T$_{c}$ [also see Fig.~1~(d)]. Therefore the phonon dispersion and linewidth measured with different $\varphi$ values can be divided into two groups, respectively. Near the nodal region, the phonon energy decreases and the linewidth increases suddenly with the same onset. In contrast, near the antinodal region, the phonon linewidth increases slowly by less than 3 meV while the energy are almost momentum-independent.

Such a dramatic behavior seen in the phonon spectra must reflect the underlying structures of the electron-phonon coupling (EPC) for this mode. In general, the momentum-dependent broadening of the phonon linewidth $\gamma_{\mathrm{EPC}}(\bm{q})$ induced by EPC can be connected to the electron polarizability $\chi(\bm{q},\Omega_{\bm{q}})$ by the EPC matrix element $g(\bm{q})$~\cite{Grimvall}:
\begin{equation}
\gamma_{\mathrm{EPC}}(\bm{q})=-|g({\bm{q})}|^{2}\texttt{Im}[\chi(\bm{q},\Omega_{\bm{q}})],
\end{equation}
where $\Omega_{\bm{q}}$ is the phonon energy. And we have assumed $g(\bm{k}, \bm{k}^\prime)=g(\bm{k}^\prime-\bm{k}) \equiv g(\bm{q})$, which is a good approximation for the particular out-of-plane apical oxygen phonon mode. The imaginary part of $\chi(\bm{q},\Omega_{\bm{q}})$ is given by:
\begin{align}
\texttt{Im}[\chi(\bm{q},\Omega_{\bm q})] & =\int_{\mathrm{BZ}}\frac{\mathrm{d}\bm{k}}{(2\pi)^{2}}\int\frac{\mathrm{d}\omega}{2\pi}A(\bm{k},\omega)A(\bm{k}+\bm{q},\omega+\Omega_{\bm{q}})\nonumber \\
 \nonumber
 & \times\left[f(\omega+\Omega_{\bm{q}}) - f(\omega)\right]\\
 & \approx\int_{\mathrm{BZ}}\frac{\mathrm{d}\bm{k}}{(2\pi)^{2}}\frac{\eta\left[f(\epsilon_{\bm{k}+\bm{q}}) - f(\epsilon_{\bm{k}})\right]}{(\epsilon_{\bm{k}}-\epsilon_{\bm{k}+\bm{q}}+\hbar\Omega_{\bm{q}})^{2}+\eta^{2}},
\end{align}
where $A(\bm{k},\omega)$ is the spectral function of quasi-electron, $f(\omega)$ is the Fermi distribution function, and $\epsilon_{\bm{k}}$ is the energy dispersion of the quasi-electrons. We have further simplified the expression in the second line of Eq. 2 by assuming that the spectral function of the quasi-electron has the form of $A(\bm{k},\omega)\approx(\eta/2)/[(\omega-\epsilon_{\bm{k}})^{2}+(\eta/2)^{2}]$, with a parameter $\eta$ characterizing the life-time of the quasi-electron due to the many body interactions.  The quasi-particle dispersion $\epsilon_{\bm{k}}$  of the BSCCO system has been measured previously (ARPES) and fitted in a tight-binding phenomenological model~\cite{RMP75-473,PRB52-615}. Im[$\chi(\bm{q},\Omega_{\bm{q}})$] can then be calculated numerically.  Fig.~3~(a) shows the calculated Im[$\chi(\bm{q},\Omega_{\bm{q}})$], which displays rather complicated features originated from the highly anisotropic electronic structure.

In addition to $\gamma_{EPC}$, the experimentally observed linewidth also include a momentum-independent background $\gamma_{0}$, which is most likely originated from surface roughness of the cleaved sample and the instrumentation broadening:
\begin{equation}
\gamma_\mathrm{exp}(\bm{q}) = \gamma_{0} + \gamma_{\mathrm{EPC}}(\bm{q}).
\end{equation}
We estimate such a background using the value of the experimental data near $\bm{q}=0$: $\gamma_0=\gamma_{\mathrm{exp}}(\bm{q}\rightarrow 0)$, since there should have been no EPC without momentum transfer. With the background subtracted, the EPC matrix element can then be determined straightforwardly with Eq.~1-3. We note that only the electron transition satisfying both energy and momentum conservations, i.e., $\bm{k}^\prime - \bm{k} = \bm{q}$ and $\epsilon_{\bm{k}^\prime} - \epsilon_{\bm{k}} = \hbar\Omega_{\bm{q}}$, contributes to the EPC-induced broadening. This imposes a stringent constraint on the possible initial and final states of electron for a given phonon momentum $\bm{q}$, as shown in Fig.~3~(c). Therefore, the determined $|g(\bm q)|^2$ should be considered as a subset of the more general $|g(\bm k, \bm k^\prime)|^2$.

\begin{figure}
\includegraphics[clip,width=3.4in]{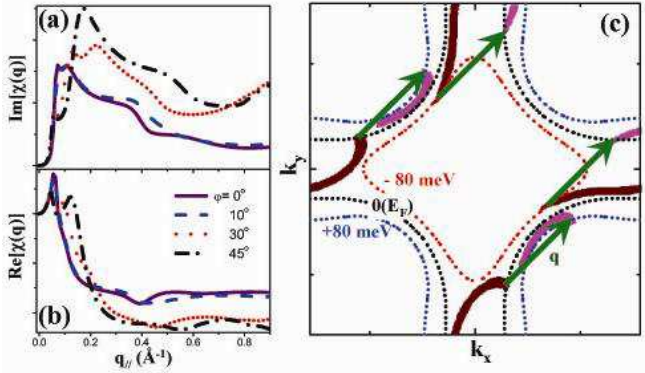}
\caption{(color online) (a) and (b) The imaginary and real parts of $\chi(\bm{q},\Omega_{\bm{q}})$ calculated with Eq.~2 at 60 K with $\eta = 10$ meV (not a sensitive parameter) for different $\varphi$, respectively. (c) The energy contour plot (dotted lines) of the quasi-particle band near E$_{F}$ of BSCCO. The initial (brown areas) states contributing to the EPC-induced broadening for q$_{\parallel}\sim$ 0.47 \AA$^{-1}$ at $\varphi=45^o$ can be determined by superimposing the density plot of the integrating function in Eq.~2 onto (c). Connecting by $\bm{q}$ (indicated by the arrows), the final states (pink areas) correspond to an energy gain that equals to $\hbar\Omega_{\bm{q}}$.}
\end{figure}

The determined EPC matrix element $|g(\bm q)|^2$ are shown in Fig.~4~(a) along different directions. With the least square method, the data can be well fitted by:

\begin{equation}
|g_{\bm{q}}|^{2}=[(2a+b)-a(\cos q_{x}+\cos q_{y})-b\cos q_{x} \cos q_{y}]^{2},
\end{equation}
which is consistent with an $A_1$ phonon mode that couples to electrons in the lattice with $C_{4v}$ symmetry. Here $a$ and $b$ are constants, representing the coupling strength between an atom displacement and its nearest and next-nearest neighboring sites, respectively. It can be seen that the experimental data, in particular, the dramatically different behaviors for nodal and anti-nodal directions, are quantitatively reproduced by our model.

\begin{figure}
\includegraphics[clip,width=3.4in]{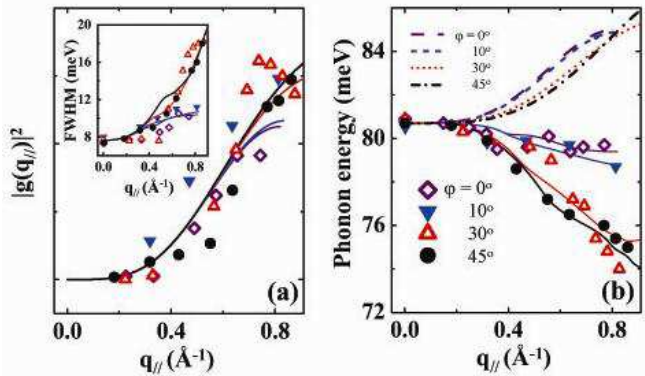}
\caption{(color online) (a) $|g_{\bm{q}}|^{2}$ determined by Eq.~1-3 and the fitting (solid lines) to Eq.~4 with $a = 0.6668$ and $b = 0.3949$. The inset shows the measured phonon linewidth as well as the data calculated with $|g_{\bm{q}}|^{2}$ for different $\varphi$. (b) The measured phonon energy with the data calculated with Eq.~5. Here we have assume a bare phonon dispersion $\Omega^{\mathrm{bare}}=83.4-1.65(\cos q_x+\cos q_y)-0.51\cos q_x\cos q_y+0.55(\cos 2q_x+\cos 2q_y)$, as plotted in broken lines.}
\end{figure}

To further test our analysis, we calculate the phonon softening induced by EPC:
\begin{align}
 &
\Delta \Omega(\bm{q})=-|g_{\bm{q}}|^{2}\texttt{Re}[\chi(\bm{q},\Omega_{\bm{q}})] \\
 & \texttt{Re}[\chi]=\int_{\mathrm{BZ}}\frac{\mathrm{d}\bm{k}}{(2\pi)^{2}}\frac{(\epsilon_{\bm{k}}-\epsilon_{\bm{k}+\bm{q}}+\hbar\Omega_{\bm{q}})\left [f(\epsilon_{\bm{k}+\bm{q}}) - f(\epsilon_{\bm{k}})\right]}{(\epsilon_{\bm{k}}-\epsilon_{\bm{k}+\bm{q}}+\hbar\Omega_{\bm{q}})^{2}+\eta^{2}} \nonumber
\end{align}
where $\texttt{Re}[\chi(\bm{q})]$ is the real part of electron polarizability, as plotted in Fig.~3~(b). Figure~4~(b) shows the experimentally observed phonon dispersion that can be reproduced by assuming a bare phonon dispersion $\Omega^{\mathrm{bare}}(\bm q)$ as the experimental background.  In particular,  the anisotropy of the phonon softening is reproduced: when $\varphi \sim 45^{o}$, the phonon energy softens dramatically at the onset q$_{\parallel} \sim$ 0.4 \AA $^{-1}$, while such a softening gets much weaker when $\varphi \sim 0^{o}$. In Fig.~4~(b), $\Omega^{\mathrm{bare}}(\bm q)$ represents the phonon dispersion if the EPC is turned off in our model, which is almost isotropic.

In conclusion, the measured phonon structure combined with a detailed knowledge of the electronic structure from ARPES can be used to determine the EPC matrix element $|g(\bm q)|^2$ directly. Our approach is completely general and could be applicable for other complex systems with highly anisotropic electronic structures. Our study also highlight the necessity of probing bosons for revealing the full structure of EBC in these complex systems.

\begin{acknowledgments}
This work is supported by China NSF-10704084, \textquotedblleft 973\textquotedblright program of China MOST (No. 2006CB921300 and No. 2007CB936800), NSF (DMR-0346826, DMR-0451163 and DMS \& E). Zhang and Guo gratefully acknowledge the support of K. C. Wong Education Foundation, Hong Kong.
\end{acknowledgments}


\begin{thebibliography}{}
\bibitem{note1} The high-energy shoulder of F$_{2}$ can be deconvoluted from the main feature by fitting the raw spectra with two standard spectral functions. The deconvolution makes no qualitative difference to the q$_{\parallel}$-dependent energy or linewidth from the raw data. We presented the deconvoluted results in this letter.
\end{thebibliography}
\end{document}